\documentclass[twocolumn]{webofc}
\usepackage[varg]{txfonts}   
\usepackage{gensymb}
\usepackage{color}

\begin{document}
\title{NoMoS: An $R \times B$ Drift Momentum Spectrometer for Beta Decay Studies}
%

\author{
	\firstname{Daniel} \lastname{Moser}\inst{1, 2 }\fnsep							\thanks{\email{daniel.moser@oeaw.ac.at}} \and 
	\firstname{Hartmut} \lastname{Abele}\inst{2} \and              
	\firstname{Joachim} \lastname{Bosina}\inst{1,3} \and
	\firstname{Harald} \lastname{Fillunger}\inst{1} \and
	\firstname{Torsten} \lastname{Soldner}\inst{3} \and
	\firstname{Xiangzun} \lastname{Wang}\inst{2} \and
	\firstname{Johann} \lastname{Zmeskal}\inst{1} \and
	\firstname{Gertrud} \lastname{Konrad}\inst{1,2}\fnsep
	\thanks{\email{gertrud.konrad@oeaw.ac.at}}
}

\institute{Stefan Meyer Institute, Austrian Academy of Sciences, Boltzmanngasse 3, 1090 Wien, Austria 
\and
           Atominstitut, TU Wien, Stadionallee 2, 1020 Wien, Austria
\and
			Institute Laue Langevin, 71 avenue des Martyrs, 38042 Grenoble Cedex 9, France
          }

\abstract{%
 The beta decay of the free neutron provides several probes to test the Standard Model of particle physics as well as to search for extensions thereof. Hence, multiple experiments investigating the decay have already been performed, are under way or are being prepared. These measure the mean lifetime, angular correlation coefficients or various spectra of the charged decay products (proton and electron). NoMoS, the \underline{n}eutron decay pr\underline{o}ducts \underline{mo}mentum \underline{s}pectrometer, presents a novel method of momentum spectroscopy: it utilizes the $R \times B$ drift effect to disperse charged particles dependent on their momentum in an uniformly curved magnetic field. This spectrometer is designed to precisely measure momentum spectra and angular correlation coefficients in free neutron beta decay to test the Standard Model and to search for new physics beyond. With NoMoS, we aim to measure inter alia the electron-antineutrino correlation coefficient $a$ and the Fierz interference term $b$ with an ultimate precision of $\Delta a/a < 0.3\%$ and $\Delta b < 10^{-3}$ respectively. In this paper, we present the measurement principles, discuss measurement uncertainties and systematics, and give a status update.
}
\maketitle

\section{Introduction}
\label{sec:intro}

The Standard Model of particle physics (SM) is the basis of our current understanding of elementary particles and their fundamental interactions. Although it describes a wide variety of phenomena and gives insights into various aspects of particle physics, current observations show its limitations (dark matter, baryon asymmetry, etc.). A very sensitive test of the SM or new physics beyond is investigating the unitarity of the Cabibbo-Kobayashi-Maskawa (CKM) matrix. If one considers the first row of the CKM matrix, $V_{\text{ud}}$ gives the most dominant contribution to the unitarity condition. There are several measurement options to determine $V_{\text{ud}}$, including superallowed ($0^+ \rightarrow 0^+$) nuclear, neutron, nuclear ($T=1/2$) mirror and pion beta decays \cite{HardyTowner15}.  The superallowed nuclear beta decays currently give the most stringent constraint as the other options' uncertainties are dominated experimentally. Recently however, the inner radiative correction $\Delta_R^V$ has been updated, resulting in a downward shift of $V_{\text{ud}}$ extracted from superallowed beta decays and a 4 $\sigma$ deviation from CKM unitarity \citep{SengVud18}, which strongly increases  the motivation for further investigations. Note that deviations from CKM unitarity can be used to perform indirect searches for physics beyond, e.g., for scalars or supersymmetry. Neutron beta decay presents a compelling alternative to determine $V_{\text{ud}}$ as it doesn't require nuclear corrections, in contrast to the superallowed decays.

\subsection*{Neutron Beta Decay in the Standard Model}

The beta decay of the free neutron is well described within the V-A theory of the SM. Fermi's Golden Rule for the neutron's decay rate yields the following insightful correlation between the mean lifetime $\tau_{\text{n}}$, the weak axial-vector coupling constant $g_\text{A}$ and $V_\text{ud}$ \cite{Czarnecki04}:
\begin{equation}
\frac{1}{\tau_{\text{n}}}=\frac{G_{\mu}^2 \vert V_ {\text{ud}}\vert ^2}{2\pi^3}m_e^5 \left(1+3g_A^2 \right)(1+\delta_R)(1+\Delta_R^V)f
\end{equation}
\label{equ:Vud}with the Fermi coupling constant $G_\mu$, the electron's mass $m_e$ as well as the outer and inner radiative corrections $\delta_R$ and $\Delta_R^V$ respectively, and the phase space factor $f$. Hence, $V_{\text{ud}}$ can be determined from independent measurements of $\tau_{\text{n}}$ and the ratio of axial-vector to vector coupling constant $\lambda = g_{\text{A}}/g_{\text{V}}$ in neutron beta decay (the conserved vector current hypothesis requires $g_{\text{V}}=1$ for zero momentum transfer). As discussed in \cite{Marciano18}, the current discrepancies in the determination of $\tau_{\text{n}}$ (significant difference between single measurement techniques) and $\lambda$ (time-dependent trend) present additional considerable motivation to further investigate this decay.

Up until now, $\lambda$ is determined most precisely from measurements of the electron asymmetry parameter $A$ \cite{Bopp86, Yero97, Liaud97, Mund13, BrownUCNA18, Markisch:2018ndu} (current Particle Data Group accuracy $\Delta A /A=0.84 \,\%$ \cite{PDG18}, which doesn't include the most recent results). Measurements of the electron-antineutrino angular correlation coefficient $a$ have not reached sub-percent accuracy yet \cite{Stratowa78, Byrne02, Darius17} (currently $\Delta a /a=2.6 \,\%$ \cite{PDG18}). However,
they offer an independent approach with significant potential for improvement. Therefore, a number of experiments is currently putting effort into improving on it \cite{zimmer00, glueck05, WIETFELDT2005, Darius17, Nab09, Baessler13, Fry18}.

The data analysis of the aSPECT experiment is almost finished and will lead to a final uncertainty of $\Delta a/a  \sim 1\% $ \citep{Guardia19}. aCORN continues data taking with an expected ultimate uncertainty of $\sim 1 \%$ \citep{Darius17}. The Nab experiment aims to determine $a$ with an ultimate precision of $  \sim 0.1 \%$ \citep{Fry18}. The coefficient $a$ can inter alia be determined from the proton momentum spectrum, which we plan to measure with NoMoS for a systematically independent determination of $\lambda$. We aim to measure $a$ with an ultimate precision of $\Delta a/a  < 0.3 \%$.

\subsection*{Probing New Physics in Neutron Beta Decay}

\begin{figure}
\centering
\includegraphics[width=0.48\textwidth]{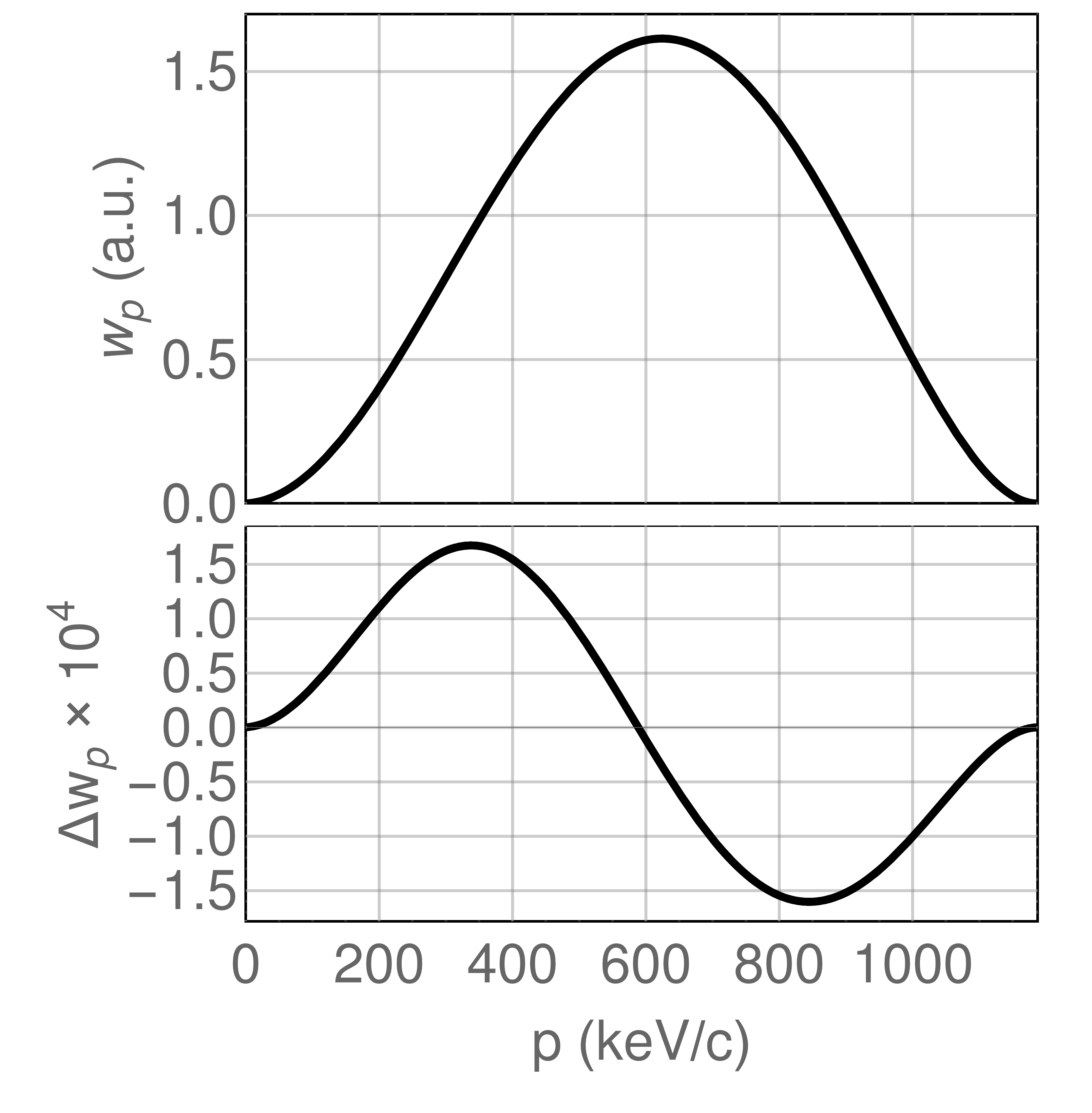}
\caption{\textbf{Top:} The electron momentum spectrum in free neutron decay for the SM value of $b=0$. \textbf{Bottom:} Deviation for other value of $b=0.001$.}
\label{fig:spectrum}
\end{figure}

In the weak interaction, extensions of the SM introduce additional couplings in beta decay \cite{jackson:1957a}, for example scalar and tensor couplings (see \cite{Herczeg01, Profumo07, Yamanaka10} for extensive discussions of possible extensions). One observable with exceptional sensitivity to these exotic couplings is the Fierz interference term $b$. A non-zero measurement of the Fierz term with $10^{-3}$ absolute sensitivity is complementary to and competitive with searches for non-SM scalar and tensor couplings in pion and muon decay and with the LHC at full luminosity and energy \citep{Baessler14, Gupta18, GONZALEZ19}. The Fierz term can be measured in a variety of beta decays with different sensitivities to scalar and tensor couplings. Despite their extraordinary precision, pure Fermi $0^+ \rightarrow 0^+$ decays are only sensitive to scalar couplings. Neutron beta decay is a mixed Fermi/Gamow-Teller transition and therefore sensitive to both scalar and tensor couplings, which further motivates a precise measurement of $b$ in neutron beta decay. Recently, the UCNA collaboration  extracted the Fierz term for the first time in neutron beta decay from existing beta asymmetry data ($b=0.067 \pm 0.005_\text{stat} \, _{-0.061}^{+0.090} \, _\text{sys}$) \cite{Hickerson17}. Note that the result's error is dominated by systematics. The \textit{Nab} experiment aims to determine $b$ with an ultimate accuracy of $\Delta b<10^{-3}$ \citep{Broussard2018}. With NoMoS we plan to measure the Fierz term with an ultimate accuracy of $\Delta b < 10^{-3}$ via the electron's momentum spectrum. Figure \ref{fig:spectrum} shows the sensitivity of the electron momentum spectrum on $b$. 

The further physics goals of NoMoS are presented in Ref. \cite{Konrad15}. Altogether, NoMoS is a promising tool to both test the SM and search for new physics beyond. 

\section{The $R \times B$ Spectrometer}
\label{sec:spectro}

As introduced in Sec. \ref{sec:intro}, we present a novel technique of momentum spectroscopy which poses an independent approach for precision measurements in neutron beta decay. NoMoS uses the $R \times B$ drift effect to separate charged particles according to their momentum. The drift velocity is given by \cite{JacksonEdyn, Plasma05}:

\begin{equation}
\label{equ:driftv}
\vec{v}_{\text{RxB}}=\frac{p}{qRB}v_{\parallel} f(\theta)\frac{\vec{R}\times \vec{B}}{RB}+ \frac{\gamma m}{q B^2}\left(\dot{\vec{v}}_{\text{RxB}} \times \vec{B}\right)
\end{equation}with the particle's momentum $p$ and charge $q$, the curvature radius $\vec{R}$ and the magnetic field $\vec{B}$, the velocity component parallel to the $B$-field $v_\parallel$, the relativistic factor $\gamma$, the particle's mass $m$ and the factor  \mbox{$f(\theta)=\left(\cos(\theta)+1/\cos(\theta)\right)/2$,} which depends on the particle's incident angle $\theta$. 

Equation (\ref{equ:driftv}) is implicit as the inertia drift (second summand) includes the time derivative of the drift velocity. The inertia drift introduces higher order contributions, though the next order is already suppressed by $\approx 10^{-3}$. Neglecting the inertia drift and assuming a constant curvature radius, a constant $B$-field and $\vec{B}$ always perpendicular to $\vec{R}$, one can integrate Eq. (\ref{equ:driftv}) over time to obtain the drift distance in zeroth order \cite{Wang2013}

\begin{equation}
\label{equ:driftD}
D_{\text{0}}=\frac{p \alpha}{q B} f(\theta)
\end{equation}where the angle of curvature $\alpha=v_\parallel T / R$ is used ($T$ is the time of travel during the drift). A huge advantage of this method is, that protons ($q=e$) and electrons ($q=-e$) drift in opposite directions and can therefore be measured separately.

\subsection*{Installation Sites}
\label{subsec:sites}
NoMoS is the first realization of an $R \times B$ spectrometer and can be used on the one hand as a standalone experiment with
\begin{itemize}
\item \textit{Beta emitters:} These are used for commissioning, calibration and characterization, and later to test the hypothesis of Lorentz invariance violation \cite{Diaz14}, or
\item \textit{Neutrons:} In this set-up (e.g., at the ILL), a beam of cold neutrons passes through a dedicated decay volume and the charged decay products are magnetically guided towards the RxB drift region (for details see next \textcolor{blue}{s}ection and Fig. \ref{fig:RxB}).
\end{itemize}
On the other hand, for high precision experiments, it can be coupled adiabatically to a magnetic field collecting the charged decay products from a long decay volume and afterwards filtering their incident angle using the magnetic mirror effect, as in the new facility PERC \cite{Dubbers08, PERC12, Wang18} or later at a PERC-like instrument at the proposed pulsed cold neutron beam facility ANNI \cite{Soldner18} at the ESS.

\subsection*{Spectrometer Design}

\begin{figure}
\centering
\includegraphics[width=0.47\textwidth]{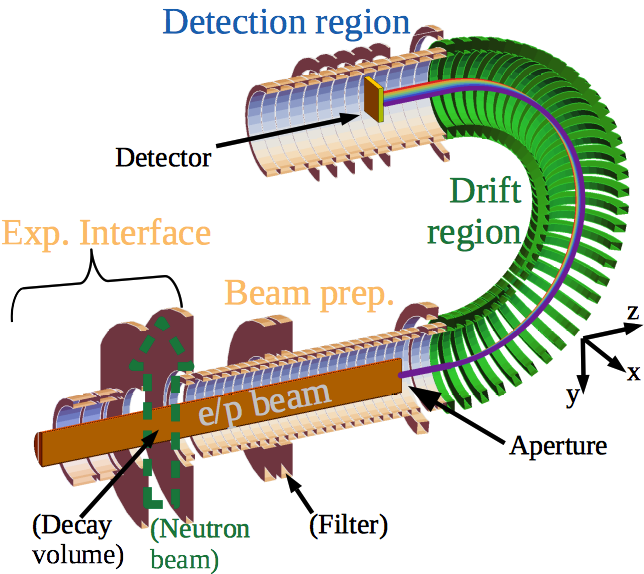}
\caption{Scheme of NoMoS: The experimental set-up is divided into four areas: the experimental interface, the beam preparation area, the drift and the detector regions. Additional features in the standalone case are given in brackets. \textbf{Particle beam:} The electron/proton beam is magnetically guided and geometrically defined by the aperture. In the drift region the particles drift according to their momentum.  The drift distance is measured in the detection region.}
\label{fig:RxB}
\end{figure}

\begin{figure*}[ht!]
\centering
\includegraphics[trim={1.1cm 20.5cm 1cm 1cm},clip,width=0.99\textwidth]{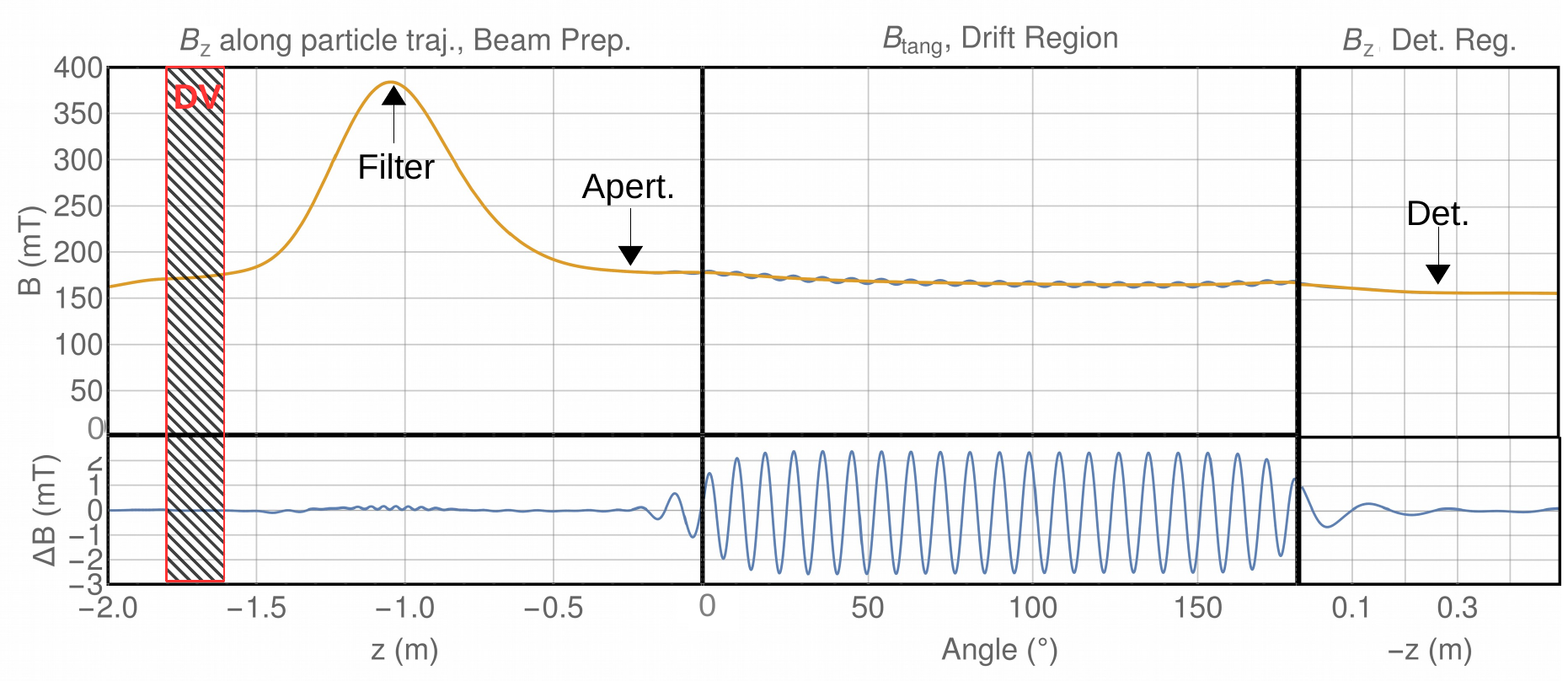}
\caption{\textbf{Top:} Shape of the magnetic field for the preliminary design of the NoMoS magnet system. Shown are the decay volume (DV - indicated with shading), beam preparation area, drift and detection regions (from left to right). Main component ($B_z$ or tangential $B_\text{tang}$) along an examplary trajectory. \textbf{Bottom:} The residuals represent the difference $\Delta B$ of the magnetic field at the particle's guiding center and its real position. The oscillation in the drift region stems from the radial gradient $\partial B_\phi / \partial R $ (for details see Sec. \ref{sec:sys}).}
\label{fig:bfield}    
\end{figure*}

NoMoS is divided into four areas: the experimental interface, the beam preparation area, the drift and the detection regions, as shown in Fig. \ref{fig:RxB}. First, cold neutrons pass through a decay volume (either in situ or external - see subsection above) while some of them decay there. Their charged decay products experience the local magnetic field $B_{\text{DV}}$ and therefore gyrate around B-field lines until they reach either the upstream end of the experiment or a magnetic filter on the downstream side. The filter has the field strength $B_\text{F}>B_{\text{DV}}$. Charged particles with an incident angle $\theta \geq \theta_{\text{max}}=\arcsin(1/\sqrt{r_\text{F}})$ with $r_\text{F}=B_\text{F}/B_{\text{DV}}$ (typical values are $r_\text{F}=2$ or $4$) are reflected from the filter by the magnetic mirror effect. The magnetically transmitted particles are then guided towards an aperture, located in the beam preparation area. The aperture defines geometrically the cross-section of the particle beam entering the drift region. It has a magnetic field $B_\text{A}<B_\text{F}$ and typical values of the magnetic field ratio $r_\text{A}=B_\text{A}/B_{\text{DV}}$ are 1 or 10. The thus prepared beam enters the drift region where  tilted $R \times B$ coils establish a B-field with constant curvature (with absolute value $B_\text{RxB} \leq B_\text{A}$, $r_\text{RxB}=B_\text{RxB}/B_{\text{DV}}$). Hence, the charged particles drift according to Eq. (\ref{equ:driftD}) with $B=B_\text{RxB}$ and $\theta=\theta_\text{RxB}=\arcsin \left( \sin{\theta_\text{DV} \sqrt{r_\text{RxB}}} \right) $. Correction coils at both ends of the drift region serve to precisely define the angle of curvature $\alpha$. After the charged particles passed the drift region, the electrons and protons are magnetically guided towards the spatial-resolving $R \times B$ drift detector. The detector is located in the detection region, which has a magnetic field of $B_\text{Det}$. If protons are measured, post-acceleration to detectable energies is required. Figure \ref{fig:bfield} shows the preliminary shape of the magnetic field through the NoMoS magnet system along an exemplary particle trajectory in the standalone case.

\subsection*{Detection and Monitoring System}

Several detectors will be installed in NoMoS, both to measure the $R \times B$ drift distance and to investigate systematic effects:

\begin{itemize}
\item \textit{$R \times B$ drift detector:} The main detector for the drift distance measurement  will have a spatial resolution of < 1 mm and ideally a surface area of $20 \times 10$ $\text{cm}^2$. Most probably, two independent detectors will be used side by side, one for electrons and one for protons.  The proton detector will be held at a high negative potential to post-accelerate them to detectable energies.  For electrons an additionally energy-resolving detector is envisaged to investigate false drift distances due to, e.g., backscattering-off of the detector itself or scattering out of the aperture (see next point).
\item \textit{Active aperture:} The aperture in the beam preparation area has a finite thickness. Through its inner face, some particles can be scattered out, potentially changing their angles and energies while still entering the drift region. This alters not only the particle distributions entering the drift region but also the drift distance distribution. To correct for this false effect, the energy-loss of the scattered particles will be measured by active surfaces at the inner face. In this way, the active aperture functions as a veto detector for false drift distances at the $R \times B$ drift detector.
\item \textit{Backscatter detector:} Some of the decay products can be backscattered-off of the $R \times B$ drift detector.  Then they deposit only part of their energy in it, and the angular distribution of the backscattered particles will range from 0 to 90$\degree$.  Hence, backscattered particles re-entering the drift region can hit the inner wall of NoMoS' vacuum vessel.  Therefore, we investigate to install an energy-resolving electron detector on the wall, along the drift tube, to detect their energy in coincidence with the $R \times B$ drift detector.
\item \textit{Beam monitor:} In the standalone case, an electron detector will be installed at the upstream end of the experimental interface to detect those electrons emitted towards the upstream end, reflected at the small magnetic field gradient in the decay volume or from the magnetic filter, and those backscattered-off of the aperture.  This detector will monitor the time stability of the particle beam and cross-check the angular selection (for details see the next section). At PERC or later ANNI, the time stability is monitored by a small parasitic monitor (out of sight of the aperture).
\end{itemize}

\section{Measurement Uncertainties and Systematics}
\label{sec:sys}

The statistical sensitivity of an $a$ or $b$ measurement with NoMoS can be determined by spectral fitting of Monte Carlo generated data. In Ref.\,\cite{glueck3}, the energy spectra have been investigated with the minimum variance bound estimator method. Table \ref{tab:stat} shows the statistical sensitivity of the proton and electron energy spectra on $a$ and $b$, respectively in comparison with the statistical uncertainty in the respective fit parameters for the drift distance spectra. Obviously, the drift distance spectra are a little less sensitive to $a$ and $b$. At the ILL and PERC (unpulsed) we expect a detection rate of about 1\,kHz.  Hence, assuming no additional fit parameters, one day of drift distance measurement yields a statistical uncertainty of $\sigma_b \approx 8 \times 10^{-4}$ and $\sigma_a /a \approx 0.28 \%$, respectively. Adding additional fit parameters accounting for systematic effects increases the required measurement time.

\begin{table}
\centering
\caption{Statistical sensitivity of the electron and proton energy spectra on $a$ and $b$ \cite{glueck3} in comparison with the drift distance spectra in the standalone case. $N$ represents the number of measured electrons or protons.}
\label{tab:stat}
\begin{tabular}{c|cc}
 & $\sigma_a$ &$\sigma_b$ \\ \hline
 energy spectrum & $2.6/\sqrt{N}$ & $7.5/\sqrt{N}$   \\
drift distance spectrum & $2.9/\sqrt{N}$ & $7.8/\sqrt{N}$ \\
\end{tabular}
\end{table}

Precision measurements of the proton and of the electron momentum spectrum require a thorough understanding of all systematic effects of the NoMoS spectrometer. We aim to understand and describe the particle transport through NoMoS and how the spectra are thereby affected as precisely as possible by  a transport function. This transport function must describe not only the $R \times B$ drift in the drift region but also the beam preparation and detection effects.  Then it can be used to fit the measured drift distance spectra. Furthermore, it can be used to investigate the sensitivity of $a$ and $b$ to systematic effects. In the following, we discuss the most important systematic effects. The investigation of these effects is still in progress, hence the numbers given in this section are preliminary.

\subsection*{Global Systematics}

\begin{itemize}
\item \textit{Magnetic field:} The magnetic field's homogeneity and absolute value have to be checked through a thorough magnetic field map as both quantities affect the transport function. The time stability will be monitored with magnetic sensors to correct for fluctuations. 
\item \textit{Adiabatic motion:} Adiabaticity should be conserved during the complete particle transport to prevent false effects in the particles' final position on the detector (and energy). This is being investigated by particle tracking simulations. Small non-adiabatic effects have to be estimated and suppressed.
\item \textit{Background:} There are several potential sources of background in NoMoS measurements. Sources for environmental background include the reactor, neighbouring experiments and cosmics. Sources for beam-related background include the collimation system and residual gas. The different contributions are disentangled by measurements with different neutron beam profiles (in the standalone case), magnetic field on/off and, for proton measurements, with post-acceleration on/off.
\item \textit{Doppler effect:} In the standalone case, cold neutrons pass through NoMoS perpendicularly to the detection system, which suppresses the Doppler effect due to neutron motion. This is not the case at PERC or later ANNI, where cold neutrons pass through NoMoS parallely to the detection system. Investigations by the PERC collaboration have shown that the mean neutron energy has to be known with a precision of better than $10^{-2}$ \cite{PERC12}.
\item \textit{Residual gas:} Proton measurements impose tight restrictions on the residual gas \cite{PERC12} (and ref. therein). Therefore, a pressure level of $10^{-9}$ mbar is desired. In addition, another neutron shutter will be implemented in order to enable automated background measurements.
\item \textit{Particle trapping:} Local magnetic field minima must be avoided as they can give rise to potential traps, especially in the decay volume. In addition, surface potential variations must be suppressed as they can lead to local field extrema and therefore give rise to potential penning traps. By spectrometer design, charged particles that gyrate around a magnetic field line in the drift region experience an oscillating magnetic field. To prevent particle trapping, the magnetic field $B_\text{RxB}$ is superimposed by a small decreasing gradient towards the detection region.
\end{itemize}

\subsection*{Beam Preparation Systematics}

All the features we use to prepare the particle beam introduce systematic effects which are being integrated into the transport function.

\begin{itemize}
\item \textit{Beam characteristics:} Potential inhomogeneities in the neutron density distribution can modify the particle spectrum entering the beam preparation area and therefore have to be determined and taken into account in the data analysis.
\item \textit{Angular selection:} Due to small inhomogeneities of the filter field ($B_F(\vec{r})$) or the decay volume field ($B_{DV}(\vec{r})$), a position dependent $r_\text{F}(\vec{r})$ is obtained, making $\theta_{\text{max}}$ position dependent. This dependency is included in the transport function and the position dependence of $r_\text{F}$ will be determined through magnetic field mapping. For a $b$ measurement on the $10^{-3}$ level, $r_\text{F}$ has to be known at the level of $\Delta r_\text{F}/ r_\text{F} \approx 10^{-3}$.
\item \textit{Edge effect:} The transmission through the aperture is position, angle and momentum dependent, which modifies the particle spectra entering the drift region. Parameters affecting this edge effect are the magnetic field ratio at the aperture $r_\text{A}$ as well as the dimensions of the aperture and its proper alignment with respect to the neutron beam, the magnetic field lines and the $R \times B$ drift detector.
\item \textit{Scattering at the aperture:} The modification of the electron spectrum due to both the scattering-off of the aperture and the scattering out through its inner face require a correction. In Ref. \cite{Dubbers08} it has been shown that the errors due to these corrections can be suppressed by making the aperture active.
\end{itemize}

\subsection*{Drift Systematics}

\begin{figure}
\centering
\includegraphics[width=0.48\textwidth]{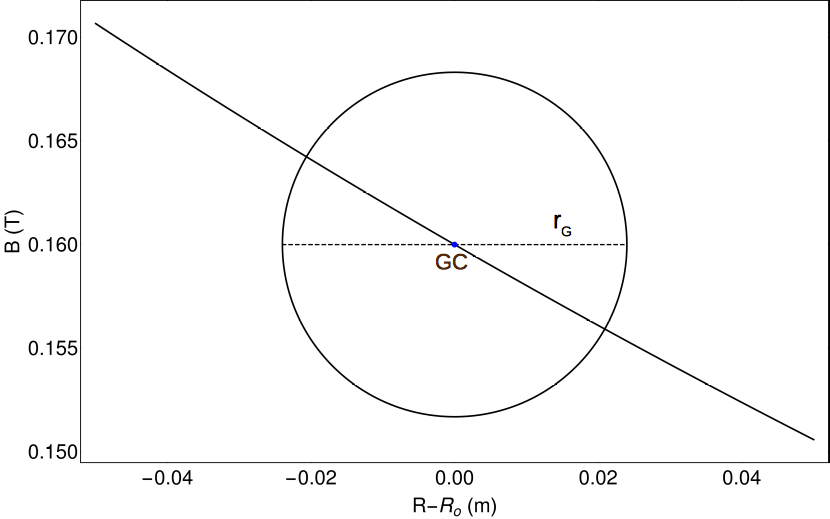}
\caption{A visualization of the effect of the radial gradient $\partial B_\phi / \partial r $ in the drift region: The main component of the magnetic field is decreasing along the radial direction. Particles which gyrate around their guiding center (GC) experience an oscillating B-field. As the gradient is not linear, the mean magnetic field is not the same as the field at the guiding center.}
\label{fig:rG}
\end{figure}

The drift systematics are defined by the magnetic field. Some of the effects can be reduced while others are unavoidable due to the design of the magnet system:

\begin{itemize}
\item \textit{Absolute $R \times B$-field:} For a $b$ measurement on the $10^{-3}$ level, the absolute $R \times B$-field value has to be known to $\Delta B_\text{RxB}/B_\text{RxB} < 10^{-4}$. Note that adding an additional fit parameter for the $R \times B$-field value decreases the systematic uncertainty by about a factor of 10 while increasing the measuring time by a factor of about four.
\item \textit{B-field gradients:} NoMoS has several B-field gradients. Some are a natural consequence of the magnet design while others are artificially introduced, mainly in order to study systematic effects:
\begin{itemize}
\item Due to the design of the $R \times B$ coils, there is an unavoidable gradient in the main component of the magnetic field, $\partial B_\phi / \partial r $ ($\phi$ is the direction along the curvature, $r$ is the radial direction). Hence,  particles passing through the drift region experience an oscillating local magnetic field. However, the gradient is not linear ($\propto 1/r$) and therefore the mean magnetic field experienced by the particles is not the same as the field at the guiding center of gyration. Figure \ref{fig:rG} shows a schematic visualization of this effect. 
\item $B_\text{RxB}$ can change over the arc length of the drift region (e.g., because of the addition of a small gradient $\partial B_\phi / \partial \phi$ along the arc length to omit magnetic traps). $r_\text{RxB}$ and the local incident angle $\theta$ are defined by this field. The effect of a small gradient is being estimated using a mean magnetic field. For a measurement of $b$ on the $10^{-3}$ level, $r_\text{RxB}$ has to be known at the level of $\Delta r_\text{RxB}/r_\text{RxB} \approx 10^{-3}$ in the drift region.
\item Along the arc length, the particles drift further and further and thereby get closer to the coils in drift direction ($x$). A very small gradient $\partial B_\phi / \partial x$ representing the increase of magnetic field towards the coils is being included in the transport function via a position dependence.
\end{itemize}
\item \textit{Opening angle $\alpha$:} The $R \times B$ drift effect acts as soon as there is a curved magnetic field. The point at which the drift gains/loses significant contribution defines the beginning/end of the curvature angle $\alpha$ in Eq. (\ref{equ:driftD}). Deviations from the nominal value $\alpha=180 \degree$ at the beginning/end $\Delta \alpha_\text{start}$ and $\Delta \alpha_\text{end}$ can be position dependent. The thorough magnetic field map will be input for particle tracking simulations, through which we will determine $\alpha$ position-dependently. Then this dependence will be integrated in the transport function. Note that adding an additional fit parameter for the opening angle $\alpha$ decreases the systematic uncertainty by a factor of about 10, while increasing the measuring time by a factor of about 2.3. Furthermore we can investigate this systematic effect by varying $\alpha$ by changing either the magnetic set-up (correction coils) or the position of the $R \times B$ drift detector (inside the $R \times B$ drift region). For a measurement of $b$ on the $10^{-3}$ level, $\alpha$ has to be known at the level of $\Delta \alpha / \alpha \approx 10^{-4}$.
\end{itemize}

\noindent 
A thorough magnetic field map is vital for the determination of most systematic effects. Hence, during commissioning of the spectrometer as well as before and after beam times, special attention will be paid to the field mapping. Hall probes will serve to measure the shape of the B-field, while NMR probes are needed for the determination of the absolute height of the $R \times B$-field.

\subsection*{Detection Systematics}

For the spatial-resolving $R \times B$ drift detector, the following systematic effects have to be taken into account: 

\begin{itemize}
\item \textit{Alignment:} A proper alignment with respect to the neutron beam, the aperture and the magnetic field lines is crucial for a quantitative drift measurement.
\item \textit{Non-active detector surface:} Normally, the area between individual detector strips ($\mathcal{O}$(10\%)) is not active. This type of binning effect is included in the transport function.
\item \textit{Detection efficiency:} The detection efficiency may vary with the particles' energy. This effect and possible corrections for it (limited fit range, calibration) have to be further investigated.
\item \textit{Backscattering:} The backscattering of decay electrons and post-accelerated protons from the detector is being investigated using scattering software simulations. For a thin detector dead layer and low detection threshold, undetected backscattering is substantially lower than the total backscattering probability. For a typical Si detector with thin Al entrance window, this probability is in the order of several percent for electrons and of one percent for protons and, for protons, its energy and angle dependence is rather small. Assuming that all backscattered particles are undetected, using additional fit parameters for the backscattering  would result in a systematic uncertainty of $\leq  1\times 10^{-4}$ on $b$ (absolute) and $< 7\times 10^{-4}$ on $a$ (relative), respectively.

\item \textit{Post-acceleration of protons:} On the one hand, the acceleration turns the protons' incident angles forward, which reduces the backscattering probability. On the other hand, the high voltage electrode can be a source of field emission and can generate additional $\vec{E} \times \vec{B}$ drift effects. Both effects have to be suppressed and are being considered in the design of the entire detection system. 
\item \textit{Edge effect:} An additional edge effect perpendicular to the drift direction is not expected as the aperture height will be chosen such that the particle beam completely fits inside the height of the detector, including two gyration radii on both sides.
\end{itemize}

\section{Summary}
\label{sec:summary}

We have presented a novel momentum spectrometer with an extensive physics program \cite{Konrad15}. Precision measurements of $a$ and $b$ are planned for tests of the SM and searches for new physics beyond. The majority of systematic effects is already included in the transport function (presented in this work), which enables a direct fit of the detected spectra. A more detailed description of the transport function is under way and accompanied by the optimization of the magnet system. In parallel, we are investigating detection systematics and are working on the design of the detection system. It is envisioned that the construction of the magnet system will start in summer 2019, after a detailed technical design study. Following a construction period of 12 to 18 months, the magnet system will be commissioned with beta emitters. A first measurement with neutrons is intended to take place at the ILL. For high precision measurements, experimental campaigns at PERC and later ANNI are planned.

\section{Acknowledgement}

We would like to thank Ferenc Glück (Karlsruhe Institute of Technology, Germany) for helpful discussions about adiabaticity, post-acceleration and the transport function, as well as Eberhard Widmann (SMI) for his support and helpful discussions. Furthermore Daniel Moser would like to thank Waleed Khalid and Raluca Jiglau (SMI) for helpful discussions.

This work is supported by the Austrian Academy of Sciences within the New Frontiers Groups Programme NFP 2013/09, the Austrian Science Fund under contracts No. W1252 (DK-PI) and P26636, the ILL under collaboration agreement No. ILL-1519.1, the TU Wien and the SMI Wien.

\bibliography{references.bib}

\end{document}